\documentclass[aip,pof, reprint]{revtex4-1}

\usepackage{amsmath}
\usepackage{graphicx}
\usepackage{upgreek}

\newcommand{\be}{\begin{equation}}
\newcommand{\ee}{\end{equation}}

\newcommand{\ba}{\begin{array}{c}}
\newcommand{\ea}{\end{array}}

\newcommand{\bd}[1]{\mathbf{#1}}
\renewcommand{\bm}[1]{\text{\boldmath $#1$}}
\newcommand{\bc}[1]{\bm{\mathcal{#1}}}

\newcommand{\rmi}{\mathrm{i}}
\newcommand{\rmd}{\mathrm{d}}
\newcommand{\rme}{\mathrm{e}}

\newcommand{\bTE}{\beta_\text{TE}}
\newcommand{\bTM}{\beta_\text{TM}}
\newcommand{\eTE}{\eta_\text{TE}}
\newcommand{\eTM}{\eta_\text{TM}}

\newcommand{\bn}{\bar{n}}
\newcommand{\bna}{\bn\alpha}
\newcommand{\tc}{\tilde{c}}
\newcommand{\td}{\tilde{d}}
\newcommand{\tA}{\tilde{A}}
\newcommand{\tB}{\tilde{B}}

\newcommand{\Pl}{P_l^1}
\newcommand{\Plp}{P_l^{1\prime}}

\newcommand{\Pkm}{P_k^{m_0}}
\newcommand{\Plm}{P_l^{m_0}}
\newcommand{\Pkmp}{P_k^{m_0\prime}}
\newcommand{\Plmp}{P_l^{m_0\prime}}

\newcommand{\half}{{\textstyle \frac1{2}}}

\newcommand{\re}{\mathrm{Re}}

\begin{document}

\title{Microdroplet oscillations during optical pulling}
\author{Simen \AA.\ Ellingsen} 
\affiliation{Department of Energy and Process Engineering, Norwegian University of Science and Technology, N-7491 Trondheim, Norway}
\email{simen.a.ellingsen@ntnu.no}

\begin{abstract}
It was recently shown theoretically that it is possible to pull a spherical dielectric body towards the source of a laser beam [Nature Photonics {\bf 5}, 531 (2011)], a result with immediate consequences to optical manipulation of small droplets. Optical pulling can be realised e.g.\ using a diffraction free Bessel beam, and is expected to be of great importance in manipulation of microscopic droplets in micro- and nanofluidics. Compared to conventional optical pushing, however, the radio of optical net force to stress acting on a droplet is much smaller, increasing the importance of oscillations. We describe the time-dependent surface deformations of a water microdroplet under optical pulling to linear order in the deformation. Shape oscillations have a lifetime in the order of microseconds for droplet radii of a few micrometers. The force density acting on the initially spherical droplet is strongly peaked near the poles on the beam axis, causing the deformations to take the form of jet-like protrusions.
\end{abstract}
\maketitle

\section{Introduction}

The use of optical forces in microfluidics is a field of research in rapid growth. Applications are already several, and many more seem to be promised\cite{monat07a,monat07b}. The use of light to manipulate microscopic flows is an attractive prospect since it is contacless and nondestructive, and easily reconfigured\cite{delville09}. This marriage of optics and fluid mechanics has been dubbed optofluidics, and was recently the topic of a focus issue of Nature Photonics\cite{nphoton}. Optical manipulation of microscopic fluid systems have been studied experimentally by several groups, demonstrating such phenomena as  manipulation of microdroplets in channels\cite{baroud07}, jet formations due to light scattering\cite{wunenburger11}, and manipulation of surfaces\cite{garnier03} and droplets\cite{verneuil09} by laser-induced Marangoni flows from surface tension gradients. Another beautiful example is the sorting of microparticles in a microfluidic environment according to size or dielectric constant using an optical lattice\cite{macdonald03}.

Optical manipulation of spherical particles is particularly attractive from an experimental point of view, not least because the force is exactly calculable without heavyweight numerics. Optical trapping\cite{ashkin86,righini07} of particles as well as pushing with radiation pressure\cite{kerker} have been the fundamental degrees of freedom in optical micromanipulation\cite{grier03,ashkinBook}. The motivation behind the present study, however, is the recent discovery that it is also possible to \emph{pull} a spherical object against the direction of propagation of a laser beam\cite{chen11}, adding new possibilities of optical manipulation. Application to spherical microdroplets is an obvious idea, yet the deformability of liquid droplets adds a new level of complexity to the theoretical description. Optical deformation of droplets was studied by Zhang and Chang some time ago\cite{zhang88}, and was the topic of a theoretical investigation by Lai et al.\ soon after\cite{lai89}, a work generalized by Brevik and Kluge\cite{brevik99}.

Chen et al.\ demonstrated\cite{chen11} that optical pulling, while impossible with a standard plane wave source, may be realized using a Bessel beam\cite{durnin87}, a type of optical beam with the additional and attractive property that it is propagation invariant and does not spread. Its usefulness for micromanipulation has already been demonstrated experimentally\cite{garches-chaves02, paterson05,milne07}. 

In this article we consider the ``optical pulling'' situation analyzed in Ref.~\onlinecite{chen11}, with a view to describing fluid motion of a droplet in the time after it has been subjected to an optical pulling pulse. The droplet, whose radius must lie in one of a few intervals a few times the wavelength of incident light, is assumed to remain spherical during the duration of the pulse, whereupon surface movement set in motion by the pulse manifests itself and eventally dies out due to viscosity. We wish furthermore to compare the pulling situation to the conventional push by a uniform plane wave, such as may be produced by a beam whose width far exceeds the extent of the droplet. 

It remains an unanswered question how large droplet deformation can be before the optical pulling effect is destroyed, however since it is an effect of subtle diffraction and interference interactions within the droplet, it is natural to presume that pulling, at least in its simplest form, requires the shape to be close to spherical. We therefore consider the case where a droplet is illuminated by a laser pulse which is short compared to the hydrodynamic response time of the droplet. This is the case studied in the classical experiment by Zhang and Chang\cite{zhang88}. The pulse transfers an impulse to the droplet, and, in addition to moving the droplet as a whole, sets surface oscillations in motion which gradually die out due to viscosity. Once oscillations have vanished, a new pulse may be transmitted, and so on.

In the next section the hydrodynamic equations of motion for the droplet surface are laid out, where perturbations are included to linear order. Section \ref{sec:force} presents the optical theory necessary to calculate the optical force density from a (general) Bessel beam as a function of polar angle acting on the initially spherical droplet, and numerical investigation of the system is considered in \ref{sec:numerics}. In this section we assume the Bessel beam to be zeroth order and consider the case of a water droplet. The optical pulling case is compared to the conventional situation of a net pushing force before conclusions in section \ref{sec:conclusions}. Various elements of the optical theory employed are found in appendices.

\section{Hydrodynamic equations of surface motion}\label{sec:motion}

We consider a droplet illuminated by a short laser pulse of duration $t_0$. As was the case in the experiment of Zhang and Chang we assume $t_0$ to be short compared to the hydrodynamical response time, so that the droplet can be approximated as being of spherical shape for the duration of the pulse. This allows the use of the well known theory of Mie scattering, as we will do in Section \ref{sec:force}. We assume the droplet to be incompressible, and may therefore omit the contribution from electrostriction from the start \cite{ellingsen11}.

\begin{figure}[tb]
  \includegraphics[width=2.5in]{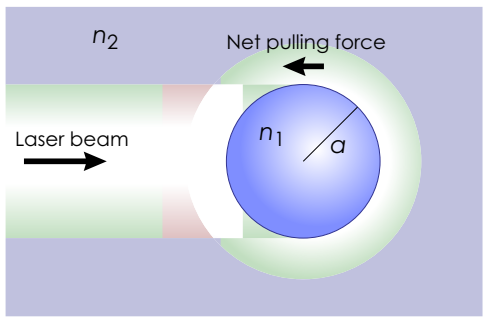}
  \caption{Sketch of the set-up considered: a droplet is illuminated by a Bessel beam such that the net force is a pull towards the laser source. A stress acts on the droplet setting in motion surface oscillations.}
  \label{geometry}
\end{figure}

A droplet irradiated by a laser beam is subjected to an optical force density distributed over the surface of the droplet. Using the completeness of Legendre polynomials, the force per unit area over the droplet surface may be written
\begin{subequations}\label{sigma}
\begin{align}
  \sigma(\Omega,t)=& \sigma(\Omega)[\Theta(t)-\Theta(t-t_0)];\\
  \sigma(\Omega) =& \sum_{l=0}^\infty\sum_{m=-l}^l\sigma_{lm}P_l^m(\cos\theta)\rme^{im\phi}.\label{sigmaexp}
\end{align}
\end{subequations}
Here $P_l^m$ are the associated Legendre functions, where in particular $P_l^0=P_l$ are Legendre polynomials, and $\Theta$ is the unit step function. $\Omega$ denotes a solid angle $(\theta,\phi)$.
We will calculate the details of this force density components $\sigma_{lm}$ in the next section, but take them to be known constants for now. The following derivation follows that of Brevik and Kluge \cite{brevik99}, in turn based on that of Lai and co-workers\cite{lai89}. For further details the reader may refer to these references.

We will find in Section \ref{sec:force} that $\sigma(\Omega)$ only depends on polar angle $\theta$, not $\phi$. Thus only the $m=0$ term contributes to the sum \eqref{sigmaexp}. Clearly, an axially symmetric force will give rise to axially symmetric surface motion, and we may simplify our notation to axial symmetry henceforth. We define
\be
  \sigma(\theta) = \sum_{l=0}^\infty\sigma_{l}P_l(\cos\theta).\label{sigmaexpl}
\ee

Let the surface of the droplet be described by the radius function
\be\label{rexp}
  r(\theta,t)=a + \sum_{l=1}^\infty h_{l}(t)P_l(\cos\theta)=a+h(t),
\ee
where $h_l(t)$ will be our unknown functions to be solved in the following. $a$ is the radius of the droplet in its inititial spherical form. Due to conservation of droplet volume, the $l=0$ solution is trivially $h_0=0$ and will not be considered further. To linear order in $h_l/a$ all the modes $l\geq 1$ automatically conserve volume since $\int_0^\pi \sin\theta P_{l\geq 1}(\cos\theta)=0$. 

We turn now to the Navier-Stokes equation governing the fluid movement in the droplet. Working to leading order in the surface perturbations, convective terms are neglected, leaving
\be\label{NS}
  \frac{\partial \bd{v}}{\partial t}=-\frac1{\varrho}\nabla p+\nu\nabla^2\bd{v}+\frac1{\varrho}\sigma(\theta,t)\delta(r-a)\hat{\bd{r}}.
\ee
Here $\varrho$ is (constant) fluid density and $\nu$ is kinematic viscosity and a hat denotes a unit vector. Incompressibility implies that $\nabla^2 p=0$ in the interior of the droplet since no electromagnetic forces act there. Thus
\be
  p(\theta)=\sum_{l=1}^\infty\Bigl(\frac{r}{a}\Bigr)^lp_{l}P_l(\cos\theta).
\ee

The viscous term in Eq.~\eqref{NS} is derived e.g.\ in Ref.~\onlinecite{brevik99} and its radial component may be written, to leading order in $h_l/a$
\begin{subequations}\label{NSvisc}
\begin{align}
  \nu\nabla^2v_r=&-2\sum_{l=1}^\infty \mu_{l}\dot h_{l}(t)P_l(\cos\theta);\\
  \mu_{l}=&\frac{\nu}{a^2}(2l^2-l-1).
\end{align}
\end{subequations}
The radial component of Eq.~\eqref{NS} in the interior now reads as $r\to a$, making use of the orthogonality of the $P_l$ expansion
\[
  \ddot h_{l}+2\mu_l\dot h_{l}=-\frac{l}{a\varrho}p_{l}.
\]

The boundary condition at the surface is that the inward pressure $\Delta p^\text{s.t.}$ minus the outward optical pressure $\sigma$ balances the hydrodynamic normal stress
at $r=a$, giving $p|_{r=a} \approx \Delta p^\text{s.t.}-\sigma$. 
The droplet's surface tension creates a pressure discontinuity across the surface whose details were worked out by Lai et al.\ \cite{lai89} to leading order in $h_l/a$:
\be
  \Delta p^\text{s.t.}=\frac{2\gamma}{a}+\frac{\gamma}{a^2}\sum_{l=1}^\infty(l^2+l-2)h_{l}(t)P_l(\cos\theta)
\ee
with $\gamma$ the surface tension coefficient.

Defining 
\be
  \omega_l^2=\frac{\gamma l}{\varrho a^3}(l^2+l-2)
\ee
we obtain the equations of motion for $h_{lm}(t)$:
\be\label{governing}
  \ddot h_{l}+2\mu_l\dot h_{l} + \omega_l^2 h_{l}=\frac{l}{\varrho a}\sigma_{l}[\Theta(t)-\Theta(t-t_0)].
\ee
This equation has the form of a forced harmonic oscillator with damping coefficient $\mu_l$ supplied by viscosity. 

Eq~\eqref{governing} can be solved quite simply using Laplace transformation, letting $\tilde h_{l}(s)=\mathcal{L}\{h_{l}(t)\}$. The Laplace transformed equation reads
\[
  \tilde h_{l}(s) = \frac{l\sigma_{l}}{\varrho a}\frac{1-\rme^{-t_0 s}}{s(s^2+2\mu_{l}s+\omega_l^2)},
\]
whereupon inverse transformation gives the solution
\begin{align}\label{motion}
&  h_{l}(t) = \frac{l\sigma_{l}}{\varrho a \omega_l^2}\Bigl\{\Bigl[1-\Bigl(\frac{\mu_l}{\gamma_l}\sin\gamma_lt+\cos\gamma_lt\Bigr)\rme^{-\mu_l t}\Bigr]\Theta(t) \notag\\
& -\Bigl[1-\Bigl(\frac{\mu_l}{\gamma_l}\sin\gamma_l(t-t_0)+\cos\gamma_l(t-t_0)\Bigr)\notag \\
&\times\rme^{-\mu_l (t-t_0)}\Bigr]\Theta(t-t_0)\Bigr\}
\end{align}
where
\[
  \gamma_l = \sqrt{\omega_l^2-\mu_l^2}.
\]
Oscillations are underdamped when $\omega_l>\mu_l$, hence $\gamma_l$ real, which is what we consider. The oscillation amplitude for mode $l$ is proportional to the force coefficient $\sigma_l$ for that mode, as one would expect. Mode $l$ has a lifetime
\[
  \tau_l=\mu_l^{-1}=\frac{a^2}{\nu(2l^2-l-1)}.
\]
For water lifetimes are in the order of a microsecond for droplets of a few micrometers' radius. Moreover we see that since $\tau\sim l^{-2}$ for large $l$, the higher order modes will damp out quickly, so the sum may be truncated at lower $l$ as time passes, which is useful. Finally we notice that when $l=1$ we have $\omega_1=\mu_1=0$, so Eq.~\eqref{governing} gives $\ddot h_1(t)=0$ for $t>t_0$. The linear behaviour of $h_1(t)$ that this implies, and the fact that $P_1(\cos\theta)=\cos\theta$, mean that the $l=1$ equation describes the translative motion of the droplet as a whole due to pulling or pushing, not a surface perturbation. For our present purposes we will not consider this term, therefore.

A note on electrostriction is warranted at this time. For an incompressible fluid the speed of sound is infinite, so that an electrostrictive force density due to the presence of the laser beam is immediately compensated by a mechanical pressure. Electrostriction therefore plays no role when it comes to the actual motion of the surface. It is of importance, however, to the droplet's stability: the inwardly directed electrostrictive force is larger than the outwardly directed net optical force. For further discussion, see \cite{brevik79,ellingsen11}.

\section{Optical force density}\label{sec:force}

With the equation of motion solution of Eq.~\eqref{motion}, what is required in order to determine the droplet surface oscillations are the coefficients $\sigma_{l}$ of the surface force density expansion \eqref{sigma}. The optical force density is obtained from Mie scattering theory \cite{mie08}. We make use of the notation of Barton et al.\cite{barton89}, to whose exposition the reader should refer for further details. In the following we denote spherical coordinates $(r,\theta,\phi)$ with $\theta$ the polar angle and origin at the droplet centre, while cylindrical coordinates are denoted $(\rho,\phi,z)$. As numerical benchmarks we have reproduced the force calculations of Refs.~\onlinecite{chen11} and \onlinecite{irvine65} for the Bessel beam and plane wave, respectively.

We consider the case of an incident field $\bc{E}^i$ impinging on an initially spherical droplet with refractive index $n_1$, embedded in a medium of refractive index $n_2$. In the present paper we shall assume $n_1$ and $n_2$ to be real for simplicity, although the generalisation to a droplet of complex $n_1$, hence absorbing droplet, is straightforward. Note that the realisation of optical pulling depends upon the imaginary part of $n_1$ being small since the droplet receives a pushing impulse for every impinging photon whose momentum it absorbs, eventually destroying the pulling effect. 

\begin{figure*}[tb]
  \includegraphics[width=5.5in]{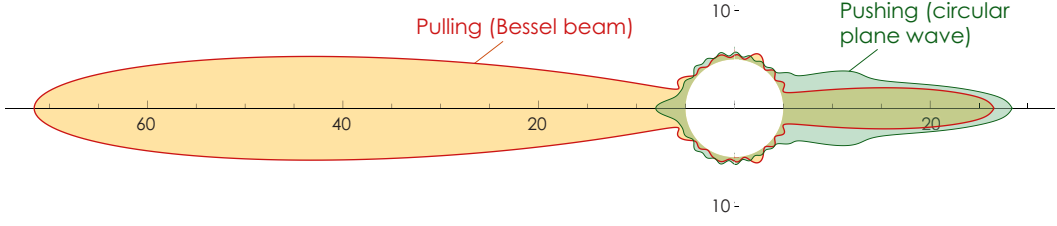}
  \caption{Polar plot of the angular distribution of force density $\sigma$ plotted as $5+\sigma(\theta)/\bar\sigma$ where  $\bar\sigma=\half\int_0^{\pi}\rmd\theta\sin\theta\sigma(\theta)$. The shading indicates the value of $\sigma(\theta)/\bar\sigma$. Comparison of pulling with order zero Bessel beam (net force towards the left) and conventional pushing by plane wave (net force towards the right). $\alpha=11.35$ in both cases. }
  \label{fig:forcedistribution}
\end{figure*}

The radial force density acting on a surface element of the sphere may be calculated by integrating the gradient of the Maxwell stress tensor
\be
  \bd{T}=\bc{E}\otimes\bc{D}+\bc{H}\otimes\bc{B}-\half(\bc{E}\cdot\bc{D}+\bc{H}\cdot\bc{B})\bd{1}
\ee
across the surface,
\be
  \sigma(\Omega)=\langle \sigma_{rr}\rangle = \langle T_{rr}(r=a^+)-T_{rr}(r=a^-)\rangle.
\ee
Here $\bd{1}$ is the unit matrix and $\langle \cdots\rangle$ denotes time average over an optical period. (For an isotropic, dielectric medium the tensors of Minkowski and Abraham coincide. The Abraham force \cite{brevik79} oscillates out and gives no contribution in the optical case.)
We let $\bc{E} = \re\{\bd{E} \rme^{i\omega t}\}$, etc., where $\bd{E},\bd{D},\bd{B}$ and $\bd{H}$ are complex field vectors. For field quantities $\bar{X}_i$ and $\bar{X}_j$, we have $\langle \bar{X}_i\bar{X}_j\rangle = \half \re\{
X_iX_j^*\}$, and in particular $\langle \bar{X}_i^2\rangle = \half |X_i|^2$.

The simplest option now is to express the stress tensor both outside and inside the spherical surface in terms of the \emph{interior} fields, since the external fields have both incident and scattered components. Using the continuity of $D_r,E_\theta,E_\phi$ and $\bd{H}$ across the surface, the force density may be written
\begin{align}\label{sigmaEM}
  \sigma(\Omega) = \frac{\varepsilon_0n_2^2}{4} (\bn^2-1)(\bn^2|E^w_r|^2+|E^w_\theta|^2+|E^w_\phi|^2).
\end{align}
Superscript $w$ signifies that the fields are evaluated just within the surface of the sphere, at radius $r=a^-$, and we have defined the shorthand
\be
  \bn = \frac{n_1}{n_2}.
\ee 

The internal fields are related to the incident electric and magnetic fields from the Bessel beam via the relations quoted in Appendix \ref{app:Mie}. For further details, refer to Refs.~\onlinecite{barton89,kerker}. The incident electromagnetic field components of an order $m$ Bessel beam are given in Appendix \ref{app:Besselbeam}. For comparison we consider a circularly polarised plane wave, the formalism for which is quoted in Appendix \ref{app:planewave}. The force density as it is distributed as a function of polar angle $\theta$ is plotted in figure \ref{fig:forcedistribution}, to which we will return for discussions later.

A Bessel beam's plane wave components form cones of angle $\theta_0$ with the direction of propagation, so that $k_z=k\cos\theta_0$ and $k_\perp=k\sin\theta_0$. A factor $\rme^{-i\omega t}$ is omitted in all field components here and henceforth.

Internal field quantities for insertion into Eq.~\eqref{sigmaEM} are calculated from the incident field quantities as laid out in appendix \ref{app:Mie}. We require the quantities $\tA_{lm}$ and $\tB_{lm}$ defined in Eq.~\eqref{AB}, which we find as
\[
  \ba\tA_{lm}\\\tB_{lm}\ea=\left\{\ba 1\\ n_2 \ea \right\}\frac{\sqrt{4\pi(2l+1)}}{l(l+1)}\sqrt{\frac{(l-m)!}{(l+m)!}}\ba A_{lm}\\B_{lm}\ea\delta_{mm_0}
\]
where we have defined
\begin{align}
&  (A,B)_{lm}=\left\{\ba i\\ 1 \ea \right\}\cdot\frac1{\psi_l(\alpha)}\int_0^1\rmd u \Bigl\{\ba \sin \alpha_z u\\ \cos \alpha_z u\ea J_m(b_\perp) u\eta_\mathrm{TM,TE}\notag \\
&\pm \ba \cos \alpha_z u\\ \sin \alpha_z u\ea\Bigl[\frac{\alpha_z}{\alpha_\perp}J'_m(b_\perp)\eta_\mathrm{TM,TE}+\frac{\rmi m \alpha}{b_\perp\alpha_\perp}J_m(b_\perp)\eta_\mathrm{TE,TM}\Bigr]\notag\\
& \times \sqrt{1-u^2}\Bigr\}P_l^m(u) \text{ for } l+m\ba \text{ even} \\  \text{ odd}\ea
\end{align}
(before and after comma pertains to $A_{lm}$ and $B_{lm}$, respectively; upper and lower pertains to even/odd values of $l+m$).
Herein, $k=2\pi/\lambda=n_2\omega/c$, $\alpha_{z,\perp} = k_{z,\perp}a$, $b_\perp = \alpha_\perp \sqrt{1-u^2}$ and 
\be\label{alpha}
  \alpha = ka=\frac{2\pi a}{\lambda_2}=\frac{n_2\omega a}{c}.
\ee
Moreover $J'_m(x) = \frac{\rmd}{\rmd x}J_m(x)$
where $J_m$ is the cylindrical Bessel function of the first kind. 
$\eTE$ and $\eTM$ are the relative weights of TE (transverse electric, w.r.t.\ direction of propagation) and TM (transverse magnetic) polarizations of the incident beam\cite{chen11}. These weights in general carry a relative phase $\delta$ between the polarizations\cite{chen11}, $\delta=\mathrm{Arg}(\eTM/\eTE)+\pi/2$.

We make the convenient definitions 
\begin{subequations}
\begin{align}
  c_{lm} =& \rmi A_{lm}[\bn\psi_l(\bna)\xi_l^{(1)\prime}(\alpha)-\psi'_l(\bna)\xi_l^{(1)}(\alpha)]^{-1}\\
  d_{lm} =& \rmi B_{lm}[\psi_l(\bna)\xi_l^{(1)\prime}(\alpha)-\bn\psi'_l(\bna)\xi_l^{(1)}(\alpha)]^{-1}
\end{align}
\end{subequations}
wherewith, using the Mie theory in appendix \ref{app:Mie}, we may finally write down the electromagnetic force density on the droplet surface from an order $m_0$ Bessel beam as
\begin{widetext}
\begin{align}\label{sigmaBig}  
  \sigma(\theta)=&\frac{n_2I_0\alpha^2}{2c}(\bn^2-1)\sum_{k,l=1}^\infty\frac{(2k+1)(2l+1)(k-m_0)!(l-m_0)!}{k(k+1)l(l+1)(k+m_0)!(l+m_0)!}\Bigl\{\frac{k(k+1)l(l+1)}{\alpha^2}c_{km_0}c_{lm_0}^*\psi_k(\bna)\psi_l(\bna)\Pkm \Plm\notag \\
  &+[c_{km_0}c_{lm_0}^*\psi_k'(\bna)\psi'_l(\bna)+d_{km_0}d_{lm_0}^*\psi_k(\bna)\psi_l(\bna)]\Bigl(\sin^2\theta\Pkmp\Plmp+m_0^2\frac{\Pkm\Plm}{\sin^2\theta}\Bigr)\notag \\
  &+m_0[c_{km_0}d_{lm_0}^*\psi_k'(\bna)\psi_l(\bna)+d_{km_0}c_{lm_0}^*\psi_k(\bna)\psi_l'(\bna)][\Pkm\Plm]'\Bigr\}
\end{align}
\end{widetext}
where we suppress the argument $\cos\theta$ of Legendre functions, $P_l^{m\prime}(x)=\rmd P_l^{m}(x)/\rmd x$, and ${}^*$ denotes complex conjugate. $I_0$ is the central intensity in the case $m_0=0$.

As promised in Section \ref{sec:motion}, $\sigma$ only depends on the polar angle $\theta$ since the incident field components depend on $\phi$ only through an overall phase factor $\exp( \rmi m_0\phi)$ which vanishes in Eq.~\eqref{sigmaEM}. This means that only $m=0$ contributes to the expansion in Eq.~\eqref{sigma}, which reduces to Eq.~\eqref{sigmaexpl}.

From definition \eqref{sigmaexpl} and orthogonality relation \eqref{Plorthog} it follows that the coefficients $\sigma_l$ are found as
\be\label{sigmal}
  \sigma_l=(l+\half)\int_0^\pi\rmd \theta \sin\theta\sigma(\theta)P_l(\cos\theta).
\ee
Formally, $\sigma_l$ requires calculation of a doubly infinite sum, yet most of the integrals resulting from inserting \eqref{sigmaBig} into \eqref{sigmal} are zero.

\section{Numerical investigation}\label{sec:numerics}

For numerical purposes, let us concentrate on the simplest case of a TM polarized Bessel beam of order $m_0=0$. This is the case considered in Fig.~1 of Ref.~\onlinecite{chen11}, and implies $\eTM=1, \eTE=0$. We let $\theta_0=78.5^\circ$. The far field scattering off a sphere from an order zero beam was recently calculated by Mitri\cite{mitri11}. 

Numerically one finds that the highest value of $l$ that must be included in expansion \eqref{rexp} is of order $\alpha$, as is physically reasonable since the wavelength of the incoming light is the smallest spatial scale of the surface force, and $\alpha$ is the number of wavelengths in a circumference. 

\subsection{Net force versus stress on the droplet}

Before considering the oscillations resulting from the optical pulling force, we will regard the integrated force acting on the front and back halves of the droplet and see how they compare to the same situation for a circularly polarized incident plane wave. This gives an intuitive picture of how the net propulsion force (obtained when integrating over the full sphere) compares with the optical stress tending to pull the droplet apart. Field equations for such a plane wave are found in Appendix \ref{app:planewave}. The force in $z$ direction acting on the front and back halves of the sphere are found by integrating over the appropriate solid angles.
Working in terms of the dimensionless force $Q$
\be\label{Qdef}
  Q = \frac{c\langle F_z\rangle}{\pi a^2 n_2 I_0};~~ \langle F_z\rangle=a^2\int \rmd \Omega \sigma(\theta)\cos\theta
\ee
half-sphere forces are ($m_0=0$, TM polarization)
\begin{align}
  Q_i =& 
  (\bn^2-1)\sum_{k,l=1}^\infty c_kc_l^*[\psi_k(\bna)\psi_l(\bna)\mathcal{P}^i_{kl}\notag \\
  &+\alpha^2\psi'_k(\bna)\psi'_l(\bna)\mathcal{R}^i_{kl}]
\end{align}
with $i=\{\text{front, back}\}$, and
\begin{align*}
  \mathcal{P}^\text{front}_{kl}=&(2k+1)(2l+1)\int_{-1}^0\rmd u P_k(u)P_l(u);\\
  \mathcal{R}^\text{front}_{kl}=&\frac{(2k+1)(2l+1)}{k(k+1)l(l+1)}\int_{-1}^0\rmd u(1-u^2) P'_k(u)P'_l(u),
\end{align*}
and corresponding quantities for the back half by replacing the integrals with $\int_0^1\rmd u$.

\begin{figure}[tb]
  \includegraphics[width=3.2in]{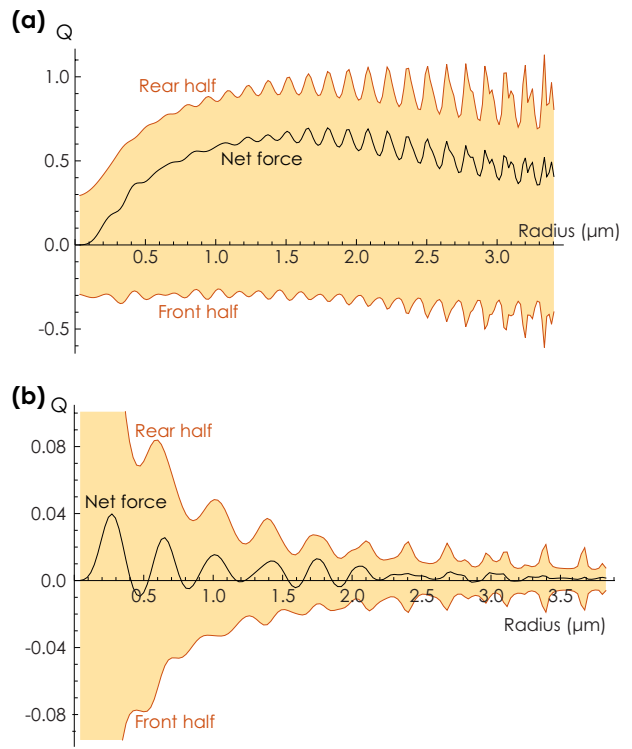}
  \caption{Nondimensionalised optical force per cross-section area, Eq.~\eqref{Qdef}, on front half and rear half of a water droplet during (a) conventional pushing by plane wave and (b) pushing/pulling by TM order zero Bessel beam. The middle graph in both panels shows the net force on the sphere. We let $\lambda=1.064\mu$m.}
  \label{fig:stress}
\end{figure}

Figure \ref{fig:stress} shows how the $z$-directed integrated surface forces acting on either half of the sphere compare with the net pulling or pushing force. Clearly a liquid droplet does not behave like two rigid hemispheres, but the graph nevertheless gives an intuitive picture of the magnitude of optical stress pulling the droplet apart as compared to the propulsion force. It is immediately clear that in regions where optical pulling is possible, i.e., where the net force in the Bessel beam case is negative, the stress is an order of magnitude larger than the net force. This is in contrast with the plane wave case where the two are in the same order of magnitude. Thus it is clear that in order to obtain a pulling force as large as the conventional pushing force, droplet oscillations set in motion by the optical stress will be comparatively greater. [Note that the difference in the numerical values of $Q$ in the two cases in Fig.~\ref{fig:stress} is of little importance; to obtain the same irradiated power, field amplitude $E_0$ must be larger for the Bessel beam than the plane wave, accounting for the difference.]

\subsection{Droplet oscillations}

For numerical purposes we take the values for water\cite{white} and let the surrounding medium be air ($n_2=1$)
\begin{align*}
  \varrho=&997\text{kg/m}^3&\gamma=0.073\text{N/m};\\
  n_1=&1.33;&\nu=1.01\mu\text{N/m}.
\end{align*}
As in Ref.~\cite{chen11} we choose $\lambda=1.064\mu$m. For optical pulling we choose the fourth pulling regime from the left in figure \ref{fig:stress}b, to wit
\[
  a=1.922\mu\text{m}; ~~~ \alpha = 11.35.
\]
The reason for not choosing the smallest $\alpha$ value is to obtain a more interesting oscillation structure which is not present when $\alpha$ is of order unity or below. Note however that the stress relative to net force is much greater for the smallest optical pulling radius interval (around $a=0.5\mu$m in figure \ref{fig:stress}b), so that oscillations will be expected to be more violent in that case, albeit somewhat compensated by the greater influence of surface tension for a smaller droplet.

\begin{figure*}[tb]
  \includegraphics[width=7in]{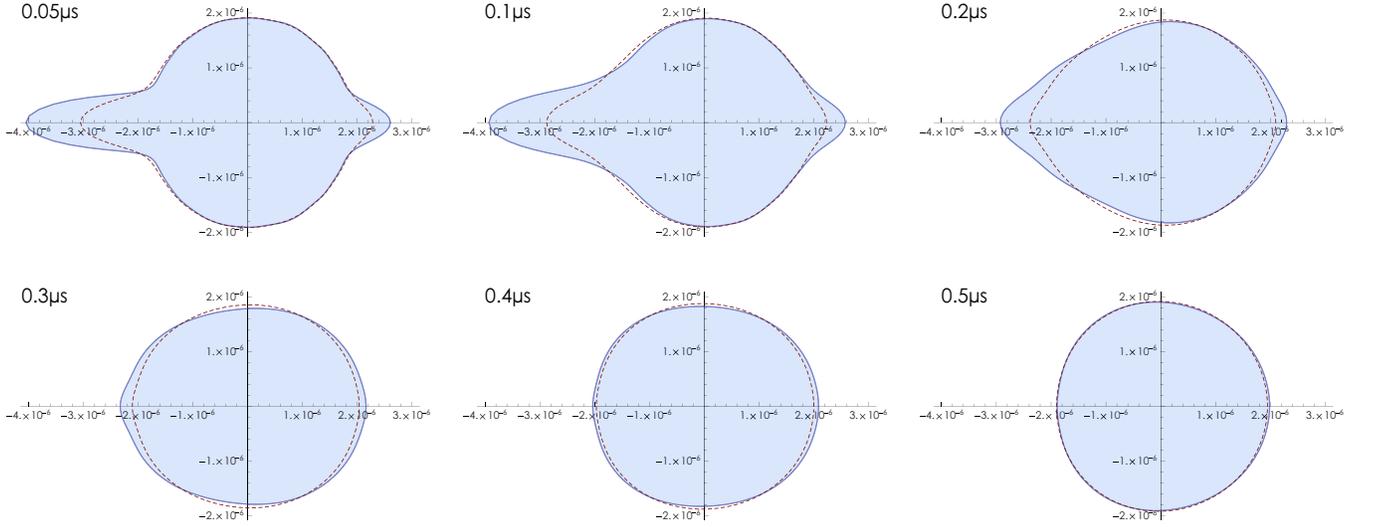}
  \caption{Droplet cross sections during optical pulling by a $m=0$ Bessel beam, pulse duration $0.04\mu$s (solid line) and $0.02\mu$s (dashed line). The laser beam enters from the left. See main text for details.}
  \label{fig:oscillations}
\end{figure*}

\begin{figure*}[tb]
  \includegraphics[width=7in]{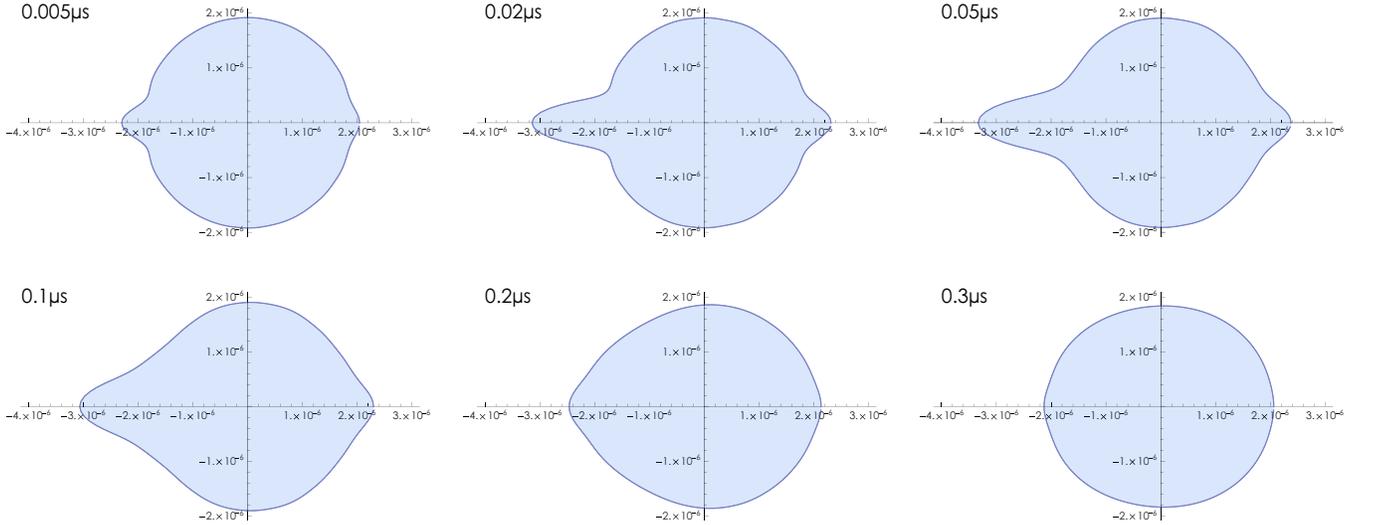}
  \caption{Same as figure \ref{fig:oscillations}, but with shorter pulse duration ($t_0=0.005\mu$s) and higher intensity ($I_0^\text{pw}=10$W/$\mu$m$^2$).}
  \label{fig:shortpulse}
\end{figure*}

In order to compare the Bessel beam pulling case with the conventional plane wave case, we let the two cases have the same laser power, i.e., the Poynting vector integrated over area $\pi a^2$ should be the same for the two cases. 
For the plane wave,  $\bd{E}^\text{pw}=E_0^\text{pw}\bd{\Lambda}\rme^{ikz}$ with $\bd{\Lambda}=\hat{\bd{x}}+\rmi \hat{\bd{y}}$. The intensities are then related by 
\be
  \frac{I_0^\text{pw}}{I_0}=\frac{\cot\theta_0}{\sin\theta_0}\Bigl[J^2_0(\alpha_\perp)-\frac{2}{\alpha_\perp}J_0(\alpha_\perp)J_1(\alpha_\perp)+J_1^2(\alpha_\perp)\Bigr].
\ee
With the numbers used in the examples, $I_0$ is then greater than $I_0^\text{pw}$ by a factor $88.2$.

For the Bessel beam we require numerical calculation of the coefficients [these are also analytically calculable, but a numerical treatment was found to be simpler]
\begin{subequations}
\begin{align}
 \mathcal{S}^1_{lmn}=&(2l+1)(2m+1)(2n+1)\int_{-1}^1\rmd u J_l(u)J_m(u)J_n(u);\\
 \mathcal{S}^2_{lmn}=&\frac{(2l+1)(2m+1)(2n+1)}{m(m+1)n(n+1)}\notag \\
 &\times\int_{-1}^1\rmd u (1-u^2)J_l(u)J'_m(u)J'_n(u).
\end{align}
\end{subequations}
Since we consider only moderate values of $\alpha$, numerical calulation of all coefficients could be done within a reasonable time.

We deliberately consider a rather high-intensity laser beam so as to visualize the oscillations most clearly.
As intensity in the plane wave case we use $I_0^\text{pw}=2$W/$\mu$m$^2$ (i.e., a laser power of ), which is about a fifth of that used in the Zhang and Chang experiment\cite{zhang88} and of that assumed by Brevik and Kluge\cite{brevik99} as well as Lai et al.\cite{lai89}. The resulting oscillations are depicted in Fig.~\ref{fig:oscillations} for different times after the onset of the laser pulse, which we assume to have duration $t_0=0.04\mu$s and $0.02\mu$s. The two different durations show the effect of reducing or lenghening pulse time. The oscillations are then somewhat larger than might reasonably be described by the linear theory herein, but nevertheless form an instructive example in that oscillations are clearly visible. The oscillatory part of the radius function \eqref{rexp} is directly proportional to $I_0$, hence the shapes of the perturbations as a function of time are the same also for smaller intensities where linear theory is accurate. The droplet has returned to approximately spherical shape after $1$-$2\mu$s. Clearly, any adverse effects of oscillations can be reduced by lowering laser intensity, at the cost of a smaller pulling force.

We find that, especially in the case of $t_0=0.04\mu$s, the droplet is deformed significantly during the duration of the pulse with our numbers. It is not known at what point deformations destroy the optical pulling effect, yet it seems likely that at this intensity level a shorter pulse might be required in practice. A similar net momentum may transferred to the droplet if one could use a shorter pulse but at higher intensity. In Fig.~\ref{fig:shortpulse} this is shown. Here, surface deformation is smaller at the the time of pulse turnoff, while the net momentum transferred (which is proportional to $I_0t_0$) is similar.

\begin{figure}[tb]
  \includegraphics[width=2.5in]{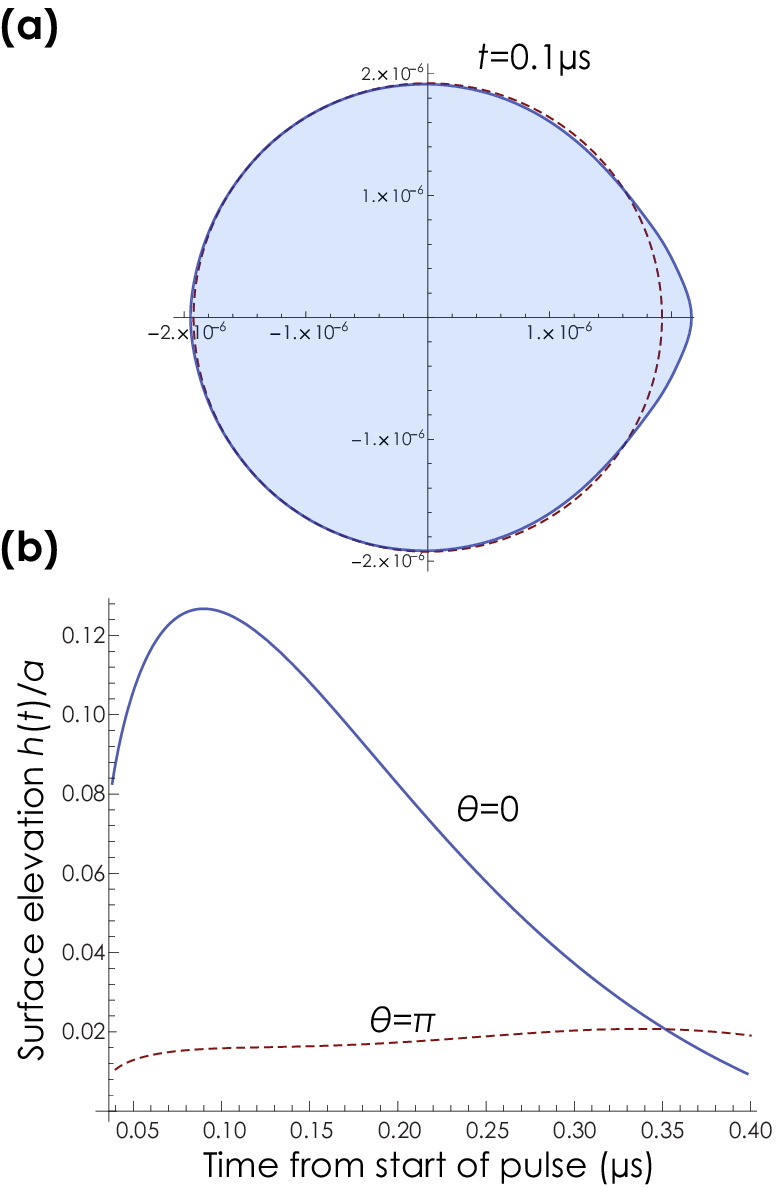}
  \caption{Surface oscillations irradiated by same power as in figure \ref{fig:oscillations}, but with a circularly polarized plane wave. Panel (a) shows approximately the maximum perturbation, panel (b) shows the relative elevation $h(t)/a$ as a function of time at back ($\theta=0$) and front ($\theta=\pi$) of droplet.}
  \label{fig:PW}
\end{figure}

In order to compare the case of optical pulling with conventional pushing we calculate the corresponding droplet oscillations when the droplet is subjected to a pulse of the same length and power, but now from a circularly polarized plane wave beam of infinite width. The formalism of such plane waves is laid out in Appendix \ref{app:planewave} (c.f.\ also Refs.~\onlinecite{lai89,jackson99,kerker}). The result is shown in figure \ref{fig:PW} for $I^\text{pw}_0=2$W/$\mu$m$^2$ and $t_0=0.04\mu$s. Although the net pushing force in this case is more than $6$ times greater than the pulling force in the corresponding case in figure \ref{fig:oscillations}, droplet deformation is much smaller. 

The reason for this is two-fold: firstly and most importantly it is due to the much greater stress-to-net force ratio in this case, as shown in Fig.~\ref{fig:stress}b, but a second effect is a relatively higher concentration of force density near the symmetry axis, which makes the initial perturbations take the form of jet-like protrusions near the droplet's poles. One may see this quite clearly in the plot of the force distribution in Fig.~\ref{fig:forcedistribution}. It is possible that the ``jet'' protrusion deformation effect can be reduced by using a higher order Bessel beam or a combination of Bessel modes; a zeroth order Bessel beam has its peak intensity near the beam center whereas higher order beams show peaks at non-zero $\rho$. We plan to investigate such possibilities in a future study.

\section{Conclusions}\label{sec:conclusions}

We have investigated the surface oscillations on a droplet under optical pulling by a zeroth order Bessel beam and made comparison to the case of pushing by a simple plane wave. The possibility to pull objects towards the source of a laser beam was predicted recently\cite{chen11}, and is expected to be of great technological interest in microfluidics. Compared to conventional pushing of a droplet, the stress acting on the droplet is much greater under pulling, setting surface oscillations in motion. Thus even though the droplets in question are small (a few micrometers' radius), surface deformations may have to be taken into account even at laser intensities where a pushing pulse results in no noticeable deformation. 

Other than the fact that droplet stress-to-net force ratio is much higher during optical pulling than pushing, additional enhancement of deformations stem from the relatively higher concentration of force density near the beam center. The force density acting on the initally spherical droplet is concentrated near the poles at $\theta=0$ and $\theta=\pi$, causing jet-like protrusions to appear along the symmetry axis. 

Relaxation times for the oscillations due to viscosity are furthermore discussed. For a water droplet of radius in the order of a few micrometers, the droplet has returned to spherical shape after $1$-$2\mu$s.

\subsection*{Acknowledgements}

The author is greatful for many discussions with and much input from Professor Iver Brevik.

\appendix

\section{Incident and internal fields from a Bessel beam}\label{app:Besselbeam}

The incident electric field $\bm{\mathcal{E}}$ from a Bessel beam of order $m_0$ is given in cylindrical coordinates $(\rho,\phi,z)$ as\cite{chen11} 
\be\label{Besselfields}
  \ba\bd{E}^i \\ \bd{H}^i\ea=\sqrt{2}\ba E_0 \\ n_2H_0\ea\rme^{\rmi k_zz+\rmi m_0\phi-\rmi\omega t}\ba\bd{e}^i \\\bd{h}^i\ea
\ee
with $H_0=E_0/(c\mu_0)$ and
\begin{subequations}
\begin{align}
  e_\rho^i =& \frac{im_0\bTE}{\rho}J_{m_0}(k_\perp \rho)+\frac{\rmi k_zk_\perp\bTM}{k}J'_{m_0}(k_\perp\rho);\\
  e_\phi^i =& -\frac{k_zm_0\bTM}{k\rho}J_{m_0}(k_\perp \rho)-\bTE k_\perp J'_{m_0}(k_\perp \rho);\\
  e_z^i =& \frac{\bTM k_\perp^2}{k}J_{m_0}(k_\perp \rho).
\end{align}
\end{subequations}
The magnetic field components are found from Maxwell's equations as
\begin{subequations}
\begin{align}
  h_\rho^i =& \frac{m_0\bTM }{\rho }J_{m_0}(k_\perp \rho)+\frac{ k_z k_\perp \bTE}{k} J'_{m_0}(k_\perp \rho);\\
  h_\phi^i =& \frac{\rmi k_z{m_0}\bTE}{k\rho}J_{m_0}(k_\perp \rho)+\rmi\bTM k_\perp J'_{m_0}(k_\perp \rho).\\
  h_z^i =& -\frac{\rmi\bTE k_\perp^2}{k}J_{m_0}(k_\perp \rho).
\end{align}
\end{subequations}
Here $\bTE = \rmi k\eTE/k_\perp^2$, $\bTM = k\eTM/k_\perp^2$. Note that $E_0$ is not the complex amplitude, but is chosen so that $\langle\bc{E}^2\rangle=E_0^2$ at $\rho=0$ for the case $m_0=0$, TM polarization. Central intensity is then $I_0=\varepsilon_0 n_2 c E_0^2$.

In spherical coordinates one readily finds
\begin{subequations}\label{BesselfieldsE}
\begin{align}
  \ba e_r^i\\e_\theta \ea=&\Bigl(\frac{\rmi {m_0}\bTE}{\rho}J_{m_0} +\frac{\rmi k_zk_\perp\bTM}{k}J_{m_0}'\Bigr)\ba\sin\theta \\\cos \theta \ea\notag \\
  &\pm\frac{\bTM k_\perp^2}{k}J_{m_0}\ba\cos\theta \\\sin\theta \ea,
\end{align}
\end{subequations}
and
\begin{subequations}\label{BesselfieldsH}
\begin{align}
  \ba h_r^i\\h_\theta \ea=&\Bigl(\frac{{m_0}\bTM}{\rho}J_{m_0} +\frac{k_zk_\perp\bTE}{k}J_{m_0}'\Bigr)\ba\sin\theta \\\cos \theta \ea\notag \\
  &\mp\frac{\rmi\bTE k_\perp^2}{k}J_{m_0}\ba\cos\theta \\\sin\theta \ea.
\end{align}
\end{subequations}
Note furthermore that $z=r\cos\theta$ in Eq~\eqref{Besselfields} and $\rho=r\sin\theta$. Component $E_\phi^i$ is naturally the same in both systems. 

\section{Mie theory relating incident and internal fields}\label{app:Mie}

The internal electric field components are expanded in spherical harmonics according to \cite{barton89}
\begin{subequations}\label{generalFields}
\begin{align}
  E_r^w=& E_0 \sum_{l=1}^\infty\sum_{m=-l}^ll(l+1)\tc_{lm}\psi_l(\bna)Y_{lm}(\Omega)\\
  E_\theta^w=& \alpha E_0\sum_{l=1}^\infty\sum_{m=-l}^l \Bigl[\bn \tc_{lm}\psi'_l(\bna)\partial_\theta Y_{lm}(\Omega)\notag \\
  &-\frac{\td_{lm}}{n_2}m\psi_l(\bna)\frac{Y_{lm}(\Omega)}{\sin\theta}\Bigr]\\
  E_\phi^w=& i\alpha E_0\sum_{l=1}^\infty\sum_{m=-l}^l \Bigl[m\bn \tc_{lm}\psi'_l(\bna)\frac{Y_{lm}(\Omega)}{\sin\theta}\notag \\
  &-\frac{\td_{lm}}{n_2}\psi_l(\bna)\partial_\theta Y_{lm}(\Omega)
\end{align}
\end{subequations}
where $\Omega=(\theta,\phi)$ and the coefficients
\begin{subequations}
\begin{align} 
  \tc_{lm} =& i\tA_{lm}[\bn^2\psi_l(\bna)\xi_l^{(1)\prime}(\alpha)-\bn\psi'_l(\bna)\xi_l^{(1)}]^{-1}\\
  \td_{lm} =& i\tB_{lm}[\psi_l(\bna)\xi_l^{(1)\prime}(\alpha)-\bn\psi'_l(\bna)\xi_l^{(1)}]^{-1}
\end{align}
\end{subequations}
and the incident field $\mathbf{E}^i$ is contained in the quantities
\be\label{AB}
  \begin{array}{c}\tA_{lm}\\ \tB_{lm}\end{array} = \frac1{l(l+1)\psi_l(\alpha)}\int\left.\begin{array}{c}E_r^i/E_0\\H_r^i/H_0\end{array}\right.Y_{lm}^*(\Omega)\rmd \Omega,
\ee
where incident fields are evaluated at $r=a^+$ and the integral is over all solid angles. 

Note that
\be
  Y_{lm}(\Omega)=\sqrt{\frac{2l+1}{4\pi}\frac{(l-m)!}{(l+m!)}}P_l^m(\cos\theta)\rme^{\rmi m\phi}
\ee
and the orthogonality relation
\be
  \int \rmd \Omega Y_{lm}(\Omega)Y_{l'm'}(\Omega)=\delta_{ll'}\delta_{mm'}
\ee
so
\begin{align}\label{Plorthog}
  \int_0^\pi \rmd \theta \sin\theta &P_l^m(\cos\theta)P_{l'}^{m}(\cos\theta)
  \frac{2}{2l+1}\frac{(l+m)!}{(l-m)!}\delta_{ll'}.
\end{align}

\section{Circularly polarized plane wave}\label{app:planewave}

A circularly polarized plane wave propagating along the $z$ direction may be expressed as (Ref.~\onlinecite{jackson99}, section 10.3)
\begin{align}
  \bd{E}^i=&E_0\bd{\Lambda}\rme^{ikz} =E_0\sum_{l=1}^\infty i^l\sqrt{4\pi(2l+1)}[j_l(kr)\bd{X}_{l1}\notag\\
  &+\frac1{k}\nabla\times j_l(kr)\bd{X}_{l1}]
\end{align}
with
\be
  \bd{X}_{lm}(\Omega)=\frac1{i\sqrt{l(l+1)}}(\bd{r}\times\nabla) Y_{lm}(\Omega)
\ee
where $k=n_2\omega/c$ and $\bd{\Lambda}=\bd{\hat{x}}+i\bd{\hat{y}}$. The radial component may then be written (Ref.~\onlinecite{jackson99}, section 10.4) at $r=a^+$
\be\label{Eir}
  E^i_r= \frac{E_0}{\alpha^2}\sum_{l=1}^\infty i^{l+1}\sqrt{4\pi l(l+1)(2l+1)}\psi_l(\alpha) Y_{l1}(\Omega).
\ee
The radial magnetic component is now found from Maxwell's equations as $H^i_r= n_2E^i_r/(\rmi\mu_0 c)$.
By using formulas \eqref{AB} and \eqref{Eir} we find
\be
  \tA_{lm} = \frac{i^{l+1}}{\alpha^2}\sqrt{\frac{4\pi(2l+1)}{l(l+1)}}\delta_{m1}=\frac{i}{n_2}\tB_{lm}
\ee
and the field components \eqref{generalFields} may be written
\begin{subequations}\label{CircPolFields}
\begin{align}
  E_r^w=&\frac{\rme^{i\phi}E_0}{\bn \alpha^2}\sum_{l=1}^\infty (2l+1)c_l\psi_l(\bna)P_l^1\\
  E_\theta^w=&-\frac{\rme^{i\phi}E_0}{\alpha}\sum_{l=1}^\infty\frac{2l+1}{l(l+1)}\Bigl[c_l\psi'_l(\bna)\Plp\sin\theta\notag \\  
  &  +d_l\psi_l(\bna)\frac{\Pl}{\sin\theta}\Bigr]\\
  E_\phi^w=&\frac{i\rme^{i\phi}E_0}{\alpha}\sum_{l=1}^\infty\frac{2l+1}{l(l+1)}\Bigl[c_l\psi'_l(\bna)\frac{\Pl}{\sin\theta}\notag \\  
    &  +d_l\psi_l(\bna)\Plp\sin\theta\Bigr]
\end{align}
\end{subequations}
having suppressed the argument $\cos\theta$ of the Legendre polynomials. Intensity is $I_0^\text{pw}=\varepsilon_0n_2 c E_0^2$.

\end{document}